\documentclass[twocolumn]{aastex631}

\defcitealias{Christensen23}{C23}

\begin{document}

\title{\bf \large Possible Contamination of the Intergalactic Medium Damping Wing in ULAS~J1342+0928 by Proximate Damped Ly\boldmath$\alpha$ Absorption}

\author[0000-0003-0821-3644]{Frederick B. Davies}
\affiliation{Max-Planck-Institut f\"{u}r Astronomie, K\"{o}nigstuhl 17, D-69117 Heidelberg, Germany}

\author[0000-0002-2931-7824]{Eduardo Ba\~{n}ados}
\affiliation{Max-Planck-Institut f\"{u}r Astronomie, K\"{o}nigstuhl 17, D-69117 Heidelberg, Germany}

\author[0000-0002-7054-4332]{Joseph F. Hennawi}
\affiliation{Department of Physics, University of California, Santa Barbara, CA 93106, USA}
\affiliation{Leiden Observatory, Leiden University, Niels Bohrweg 2, 2333 CA Leiden, Netherlands}

\author[0000-0001-8582-7012]{Sarah E. I. Bosman}
\affiliation{Institute for Theoretical Physics, Heidelberg University, Philosophenweg 12, D–69120, Heidelberg, Germany}
\affiliation{Max-Planck-Institut f\"{u}r Astronomie, K\"{o}nigstuhl 17, D-69117 Heidelberg, Germany}

\begin{abstract}
    The red damping wing from neutral hydrogen in the intergalactic medium is a smoking-gun signal of ongoing reionization. One potential contaminant of the intergalactic damping wing signal is dense gas associated with foreground galaxies, which can give rise to proximate damped Ly$\alpha$ absorbers. The Ly$\alpha$ imprint of such absorbers on background quasars is indistinguishable from the intergalactic medium within the uncertainty of the intrinsic quasar continuum, and their abundance at $z\gtrsim7$ is unknown. Here we show that the complex of low-ionization metal absorption systems recently discovered by deep JWST/NIRSpec observations in the foreground of the $z=7.54$ quasar ULAS~J1342$+$0928 can potentially reproduce the quasar's spectral profile close to rest-frame Ly$\alpha$ without invoking a substantial contribution from the intergalactic medium, but only if the absorbing gas is extremely metal-poor ($[{\rm O}/{\rm H}]\sim-3.5$).  Such a low oxygen abundance has never been observed in a damped Ly$\alpha$ absorber at any redshift, but this possibility still complicates the interpretation of the spectrum. Our analysis highlights the need for deep spectroscopy of high-redshift quasars with JWST or ELT to ``purify'' damping wing quasar samples, an exercise which is impossible for much fainter objects like galaxies.
\end{abstract}


\section{Introduction}

The detection of neutral hydrogen in the intergalactic medium (IGM) via the extended damping wing of the Ly$\alpha$ line has been sought after for many years as a smoking gun of an ongoing reionization process (e.g.~\citealt{ME98,Mortlock11,Greig17b,Banados18,Davies18b}). One important caveat of any putative IGM damping wing signal is that a very similar absorption profile can be produced by dense gas within a foreground galaxy, i.e.~a damped Ly$\alpha$ absorber (DLA; e.g. \citealt{Wolfe05}). 
Such absorption systems universally exhibit absorption lines from low-ionization metal species such as \ion{Si}{2}, \ion{O}{1}, \ion{C}{2}, and \ion{Mg}{2} (e.g. \citealt{Prochaska03,Cooke17}), so an IGM-mimicking DLA close to a quasar should then imprint these additional absorption features redward of Ly$\alpha$ (see e.g.~\citealt{Banados19,Andika22}). While sensitive searches can be performed for many of these species with the same spectroscopy used for the Ly$\alpha$ damping wing investigations, with a non-detection one can only ever place an upper limit on the metallicity of an absorber that could mimic the IGM signal. 
Nevertheless, if these limits are sufficiently stringent so as to rule out anything but a record-breaking metal-poor absorber (e.g.~\citealt{Banados18,Wang20}), the IGM hypothesis can (at least tentatively) be preferred.

With the advent of JWST/NIRSpec \citep{Jakobsen22}, the searches for these metal line systems can be performed at much higher sensitivity and without confounding atmospheric contamination. The first such survey performed on high-redshift quasars has recently been published by \citet[][henceforth \citetalias{Christensen23}]{Christensen23}. In particular, they uncovered a complex of low-ionization metal absorption systems at $z>7.3$ towards the quasar ULAS~J1342$+$0928 (henceforth J1342), a luminous $z>7.5$ quasar whose damping wing signal has been well-studied \citep{Banados18,Davies18b,Greig19,Greig22,Durovcikova20,Reiman20}.
The highest redshift absorption system found by \citetalias{Christensen23} is at $z=7.476$, only $2300$\,km/s from the systemic redshift of the quasar, and could contaminate the observed damping wing if its associated \ion{H}{1} column density is large enough.

\begin{figure*}[ht]
\begin{center}
\resizebox{8.5cm}{!}{\includegraphics[trim={0.0em 1em 0.0em 0em},clip]{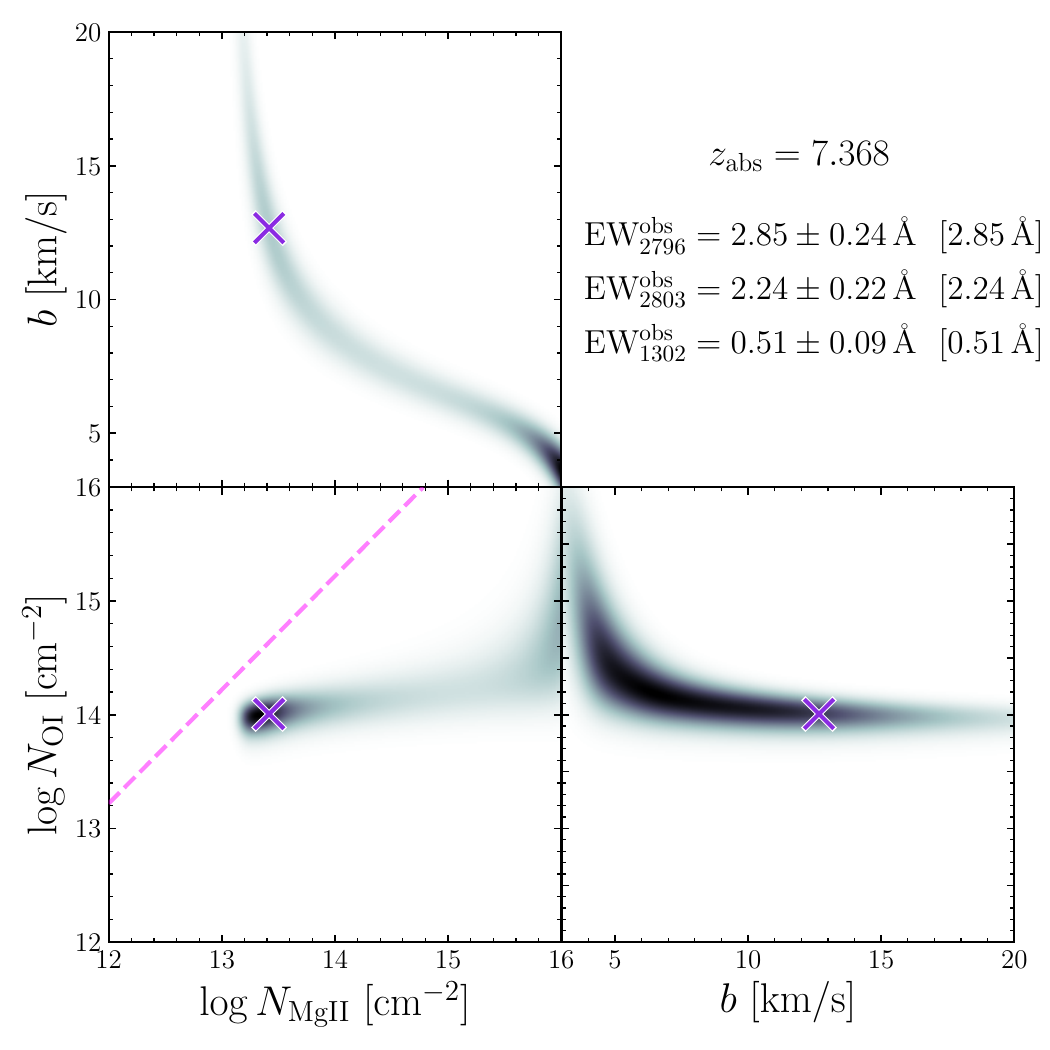}}
\resizebox{8.5cm}{!}{\includegraphics[trim={0.0em 1em 0.0em 0em},clip]{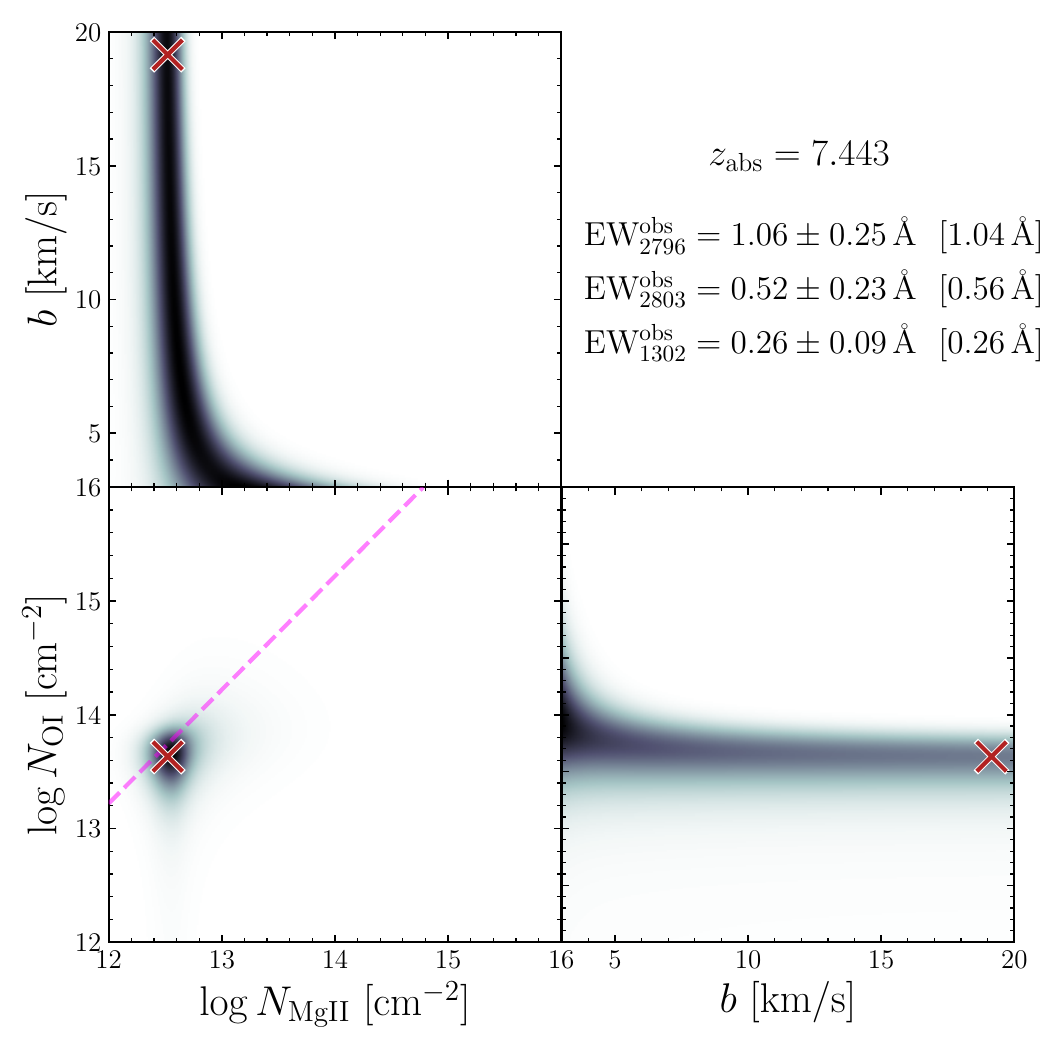}}\\
\resizebox{8.5cm}{!}{\includegraphics[trim={0.0em 1.1em 0.0em 0em},clip]{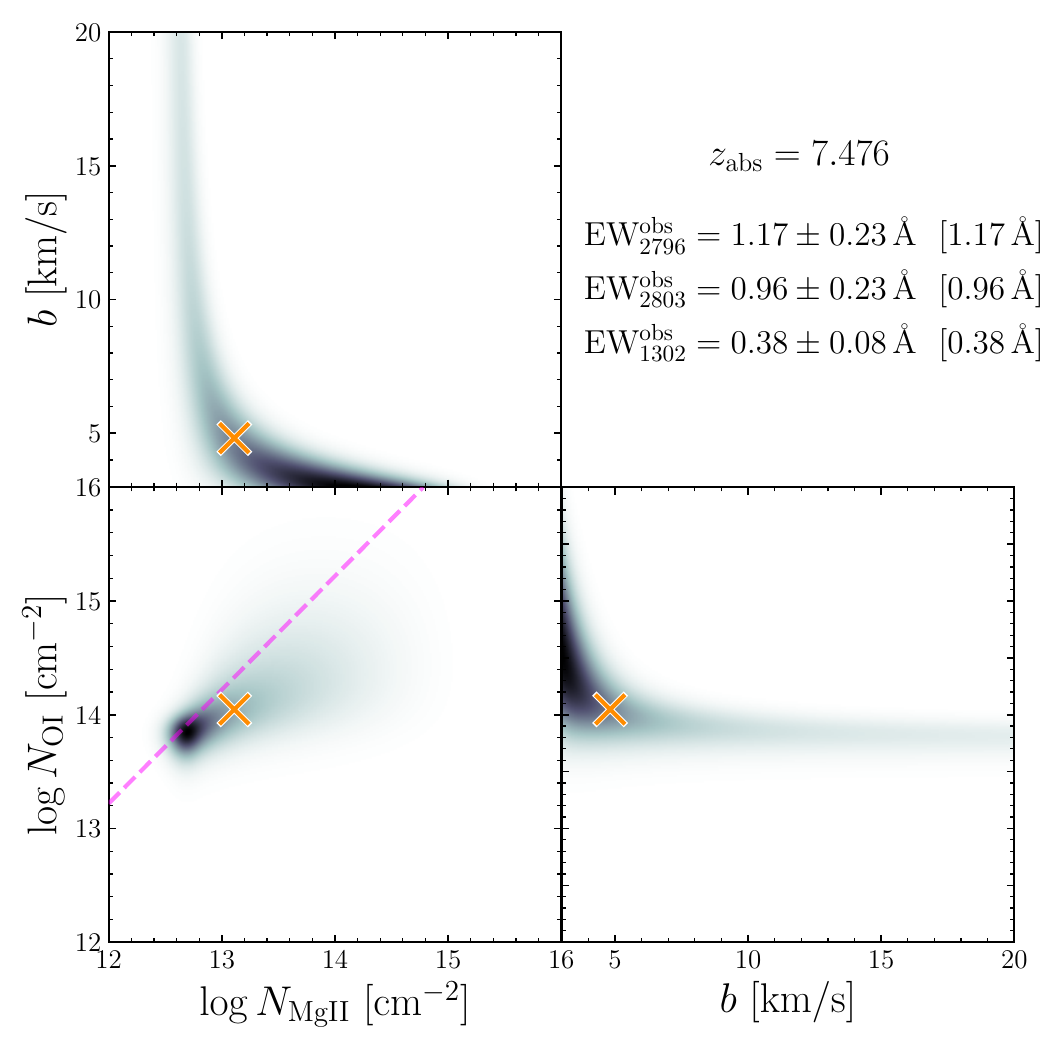}}
\end{center}
\caption{Marginalized likelihood on the \ion{Mg}{2} column density, Doppler parameter, and \ion{O}{1} column density for the $z=7.368$ (left), $z=7.443$ (right), and $z=7.476$ (bottom) absorption systems from \citet{Christensen23}. The crosses mark the maximum of the 3D likelihood that we adopt as our fiducial absorber parameters. The dashed line in the lower left sub-panels shows where the {\rm OI}/{\rm MgII} ratio would be equivalent to the solar abundance ratio of O/Mg from \citet{Magg22}. The measured \ion{Mg}{2} $\lambda\lambda2796,2803$ and \ion{O}{1} $\lambda1302$ equivalent widths reported by \citetalias{Christensen23} are shown in the upper right of each absorber panel, along with the corresponding values from our best-fit parameters in brackets.}
\label{fig:abslike}
\end{figure*}

In this letter, we explore the possibility that the \mbox{$z>7.3$} absorbers discovered by \citetalias{Christensen23} host high \ion{H}{1} column density DLAs, and investigate their potential impact on the resulting IGM neutral fraction constraint. In Section~\ref{sec:abs} we use the \ion{Mg}{2} and \ion{O}{1} equivalent widths to estimate plausible metal column densities associated with the three absorbers, and translate them into metallicity-dependent \ion{H}{1} column densities. In Section~\ref{sec:xhi}, we redo the IGM damping wing analysis of \citet{Davies18b} with the DLA systems injected in post-processing, and show how they quantitatively change the resulting constraints on the IGM neutral fraction. Finally, we conclude in Section~\ref{sec:disc} with a discussion of the implications for future studies of the IGM during the reionization epoch using quasar spectroscopy.

We assume a flat $\Lambda$CDM cosmology with $h=0.685$, $\Omega_{\rm m}=0.3$, and $\Omega_{\rm b}=0.047$. Although the most precise systemic redshift of J1342 currently available is $z_{\rm sys}=7.5400\pm0.0003$ \citep{Banados19b}, we adopt the very similar redshift from \citet{Venemans17b}, $z_{\rm sys}=7.5413\pm0.0007$, for consistency with \citealt{Davies18b}, but note that this difference ($\sim45$\,km/s) is very small relative to the velocity separation between the quasar and the foreground absorbers.

\section{A plausible model for the $z>7.3$ absorbers in J1342}\label{sec:abs}

We analyze the three $z>7.3$ absorption systems in the spectrum of J1342 identified by \citetalias{Christensen23}. If an individual absorption line is unsaturated, the corresponding column density can be directly computed from its equivalent width (EW). If the line is saturated, then the EW becomes highly sensitive to the kinematics of the absorber. The intrinsic ratio of the \ion{Mg}{2} $\lambda\lambda2796,2803$ doublet is two, i.e. in the linear regime the EW of the $\lambda2796$ line should be twice the EW of the $\lambda2803$ line. However, as described more quantitatively below, the measured ratios of the $z>7.3$ absorbers in \citetalias{Christensen23} differ from this ratio in some cases, suggesting that saturation may play a role. In addition, as noted by \citetalias{Christensen23}, while the absorption lines appear unsaturated (i.e., central flux values well above zero) in the JWST/NIRSpec data, the spectral resolution is only $R\sim2000$, which broadens the observed absorption lines by $\sim150$\,km/s. This is more than an order of magnitude broader than the typical $\sim3$--$15$\,km/s intrinsic widths of metal-poor DLAs and weak \ion{Mg}{2} absorbers at lower redshift (e.g.~\citealt{Cooke11,Cooke17,Churchill20}). We constrain the possible range of \ion{Mg}{2} and \ion{O}{1} column densities associated with the absorbers by fitting a single-component Voigt profile model to their observed EWs. 

We aim to obtain a metal column density that we can translate into a neutral hydrogen column density $N_{\rm HI}$ under a range of possible abundances. The best tracer of \ion{H}{1} is \ion{O}{1} due to the charge-exchange equilibrium imposed by their nearly identical photoionization energies (e.g.~\citealt{Draine11}). However, with only an unresolved measurement of the \ion{O}{1} $\lambda1302$ line, we have no handle on the degree of saturation. We therefore model the absorption systems by considering \ion{Mg}{2} and \ion{O}{1} jointly\footnote{Other low-ionization species are present in these systems, but some are marginally detected or blended with lower-redshift systems, so we opt for a more consistent analysis.} We fit a three-parameter single-component model for each absorber, consisting of the \ion{Mg}{2} column density $N_{\rm MgII}$, the Doppler parameter $b$, and the \ion{O}{1} column density $N_{\rm OI}$, to the EW values of \ion{Mg}{2} $\lambda\lambda2796,2803$ and \ion{O}{1} $\lambda1302$ measured by \citetalias{Christensen23}. Here we explicitly assume that \ion{Mg}{2} and \ion{O}{1} trace the same gas, which should be a reasonable assumption as they are both most abundant in the neutral gas phase.

In Figure~\ref{fig:abslike} we show the best-fit trio of parameter values (given in Table~\ref{tab:abs}) for each absorber as crosses, overlaid onto the two-dimensional marginalized likelihoods of each pair of parameters assuming flat priors of $N_{\rm MgII}=10^{12}$--$10^{16}$\,cm$^{-2}$, $N_{\rm OI}=10^{12}$--$10^{16}$\,cm$^{-2}$, and $b=3$--$20$\,km/s, where the $b$ prior is motivated by the range of observed kinematics of the most metal-poor DLA systems at lower redshift \citep{Cooke11,Cooke17}. Note that the peak in three-dimensional parameter space does not always lie within an obvious peak in the two-dimensional marginalized spaces -- we show them only for illustration. The $N_{\rm MgII}$-$b$ likelihood space (upper panels) demonstrates the effect of saturation: if the $b$ parameter is small enough, the \ion{Mg}{2} column density can be (almost) arbitrarily high. In the lower left sub-panels, the dashed line corresponds to the solar abundance ratio of O/Mg from \citet{Magg22}. We discuss the absorbers individually below, ordered by increasing redshift.\\

\noindent \emph{The $\textit{z=7.368}$ absorber:}
The best-fit parameters are $\{\log{N_{\rm MgII}/{\rm cm}^{-2}},\,b,\,\log{N_{\rm OI}/{\rm cm}^{-2}}\}=\{13.43,12.6\,{\rm km}/{\rm s},14.01\}$. This absorber exhibits a \ion{Mg}{2} doublet ratio of $1.27$, inconsistent with the intrinsic one (EW$_{2796}$/EW$_{2803}=2$), suggesting a moderate level of saturation. The best-fit \ion{O}{1} column density is a factor of four smaller than one would expect assuming a solar abundance pattern and no ionization correction. This deficit may suggest a sub-solar $[{\rm O}/{\rm Mg}]$, or that the oxygen (and therefore hydrogen) neutral fraction in the absorber is substantially below one. \\

\noindent \emph{The $\textit{z=7.443}$ absorber:}
The best-fit parameters are $\{\log{N_{\rm MgII}/{\rm cm}^{-2}},\,b,\,\log{N_{\rm OI}/{\rm cm}^{-2}}\}=\{12.52,19.1\,{\rm km}/{\rm s},13.63\}$. This absorber exhibits a \ion{Mg}{2} doublet ratio almost exactly equal to the intrinsic ratio of 2, suggesting it is not saturated, but the weaker line is only detected at just over $2\sigma$ significance. The corresponding \ion{O}{1} column density is perfectly consistent with the \ion{Mg}{2} column density assuming a solar abundance pattern.\\

\noindent \emph{The $\textit{z=7.476}$ absorber:}
The best-fit parameters are $\{\log{N_{\rm MgII}/{\rm cm}^{-2}},\,b,\,\log{N_{\rm OI}/{\rm cm}^{-2}}\}=\{13.11,4.8\,{\rm km}/{\rm s},14.05\}$. This absorber has \ion{Mg}{2} doublet lines with almost equal EWs, suggesting it is highly saturated and thus very kinematically cold. However, the uncertainties in the EWs are large enough that a ratio of two is not significantly ruled out.\\

\begin{table}
    \centering
    \begin{tabular}{c|ccc}
        $z_{\rm abs}$ & $\log{[N_{\rm MgII}/{\rm cm}^{-2}]}$ & $b$\,[km/s] & $\log{[N_{\rm OI}/{\rm cm}^{-2}]}$ \\
         \hline \hline
        7.368 & 13.43 & 12.6 & 14.01\\
        7.443 & 12.52 & 19.1 & 13.63\\
        7.476 & 13.11 & 4.8 & 14.05\\
    \end{tabular}
    \caption{Adopted physical properties of the $z>7.3$ absorbers, corresponding to the maximum likelihood in the 3D space of $\{N_{\rm MgII},b,N_{\rm OI}\}$ from the \ion{Mg}{2} $\lambda\lambda2796,2803$ and \ion{O}{1} $\lambda1302$ equivalent widths from \citet{Christensen23}.}
    \label{tab:abs}
\end{table}

We stress that the observational constraints on the absorber EWs are loose enough to allow a wide range of saturation, so any quantitative constraints one might attempt to place on the column densities are strongly dependent on the prior one sets on the Doppler parameter; the shape of the marginalized likelihoods in Figure~\ref{fig:abslike} are highly dependent on whether, for example, a linear or logarithmic prior is placed on $b$, or if $b$ is allowed to be arbitrarily small. For example, if we fix $b=1$\,km/s, the observed EWs of the $z=7.476$ absorber can still be closely reproduced (i.e. well within the observed uncertainties) with \ion{Mg}{2} and \ion{O}{1} column densities a factor of 100 higher than our fiducial model. Such high column densities are seen in lower redshift systems, but they are exceedingly rare, with an occurrence rate roughly 1000 times lower than systems with our adopted column densities (e.g.~\citealt{Churchill20}). In this work, we consider the best-fit trio of parameter values as a ``plausible'' solution to the physical properties of the absorber, and treat them as fixed for the rest of the following. We note that our best-fit column densities are either very similar to the values one would derive assuming the lines are fully unsaturated, or only modestly boosted (up to $\sim0.3$\,dex in $N_{\rm OI}$) by taking saturation into account.

\subsection{Neutral hydrogen content}

From $N_{\rm OI}$, we compute the neutral hydrogen column density $N_{\rm HI}$ by assuming an oxygen abundance relative to solar $[{\rm O}/{\rm H}]$, and the solar oxygen abundance from \citet{Magg22}\footnote{The difference from the oft-used \citet{Asplund09} oxygen abundance is only $+0.08$ dex, so this choice does not affect our results substantially.}. We choose two representative abundances: $[{\rm O}/{\rm H}]=-2.5$, representing a ``typical'' high redshift DLA metallicity extrapolated to $z>7$ from the $z<5.5$ population \citep{Rafelski12,Rafelski14,DeCia18} and similar to the very metal-poor DLA population at lower redshift (e.g.~\citealt{Cooke11}), and $[{\rm O}/{\rm H}]=-3.5$, a factor of a few lower metallicity than the lowest metallicity DLAs known at any redshift (e.g.~\citealt{Cooke17,D'Odorico18}). 

The resulting $\log{N_{\rm HI}}/{\rm cm}^{-2}$ values are \{19.74, 19.36, 19.78\} ($[{\rm O}/{\rm H}]=-2.5$) and \{20.74, 20.36, 20.78\} ($[{\rm O}/{\rm H}]=-3.5$) for the $z_{\rm abs}=\{7.368,7.443,7.476\}$ absorbers. In the higher metallicity case, none of the absorbers actually reaches the $N_{\rm HI}=10^{20.3}$\,cm$^{-2}$ threshold to be considered a DLA \citep[e.g.][]{Wolfe05}. In the lower metallicity case, all three absorbers would be DLAs with $\log{N_{\rm HI}/{\rm cm}^{-2}}\simeq20.4$--$20.8$. As shown by \citet{Wang20}, the IGM damping wing from a fully neutral IGM at $z=7$ is comparable to that from a DLA with $N_{\rm HI}=10^{21}$\,cm$^{-2}$, so the comparably high column density of the highest redshift absorber suggests that it could play a role in the observed damping wing signal, which we now discuss in more detail.

\section{Impact of proximate absorption on IGM constraints}\label{sec:xhi}

\begin{figure*}[ht]
\begin{center}
\resizebox{15cm}{!}{\includegraphics[trim={1.5em 2.0em 1.0em 1em},clip]{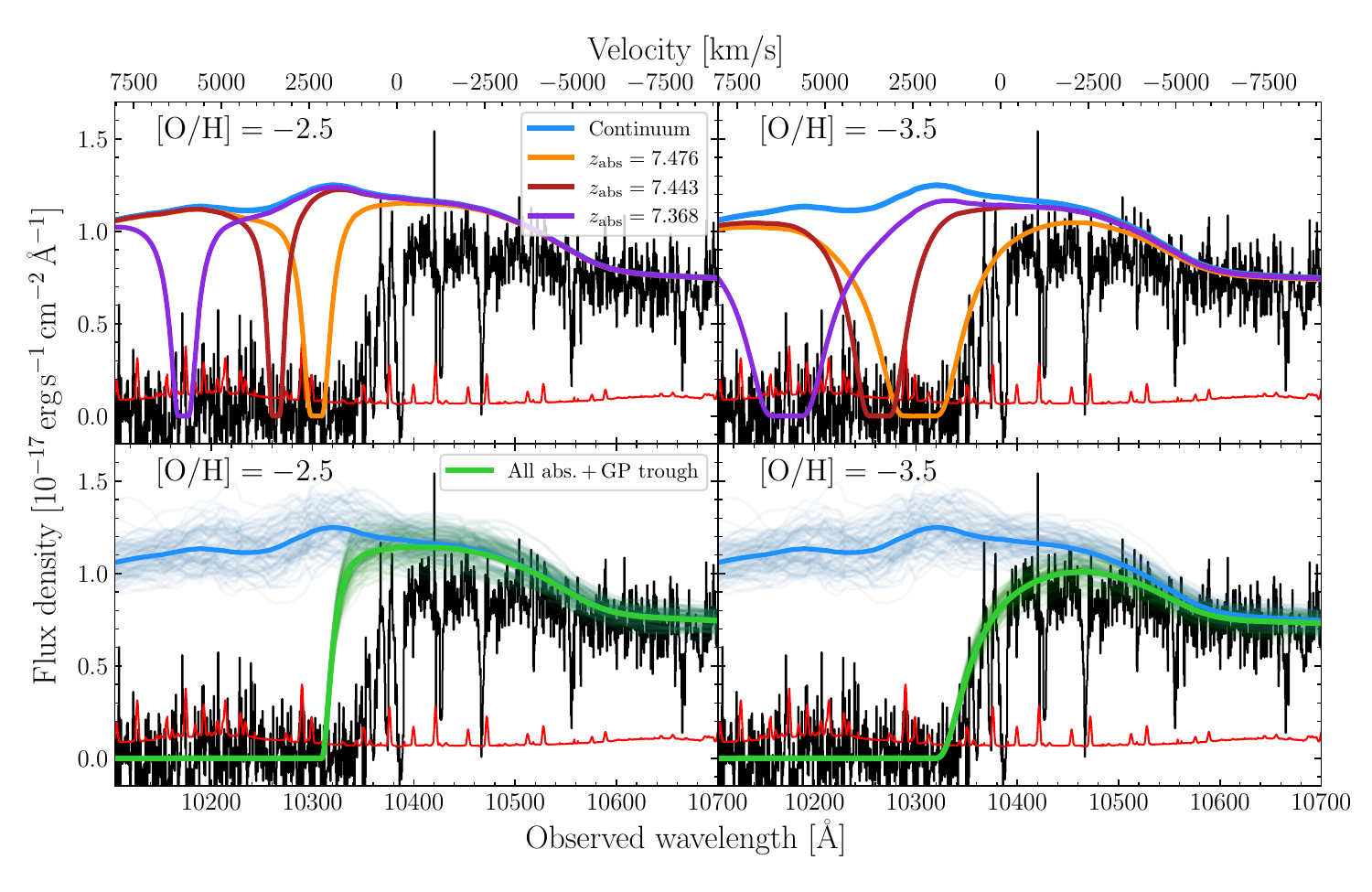}}
\end{center}
\caption{Spectrum of J1342 (black + noise vector in red) and a prediction for its unabsorbed continuum (blue). The top panels show the individual \ion{H}{1} absorbers corresponding to the three absorption systems from \citet{Christensen23}, while in the bottom panel the absorbers are combined along with a complete Gunn-Peterson trough at wavelengths shortward of Ly$\alpha$ in the frame of the highest redshift system ($z=7.476$). The left panels show the \ion{H}{1} column densities assuming an absorber metallicity of $[{\rm O}/{\rm H}]=-2.5$, while the right panels assume $[{\rm O}/{\rm H}]=-3.5$. The transparent swarm of curves in the lower panels show draws from the quasar continuum prediction uncertainty and their resulting impact after absorption.}
\label{fig:absmodel}
\end{figure*}

In Figure~\ref{fig:absmodel}, we show the Magellan/FIRE spectrum of J1342 from \citet{Banados18}, which was also used in the damping wing analysis of \citet{Davies18b}. We also show our two representative models for the \ion{H}{1} absorption corresponding to the higher metallicity ($[{\rm O}/{\rm H}]=-2.5$, left panels) and lower metallicity ($[{\rm O}/{\rm H}]=-3.5$, right panels) cases, where in all panels the blue curves (at the top) show the prediction for the unabsorbed quasar spectrum from \citet{Davies18a}. In the top panels we show the individual absorbers at $z_{\rm abs}=\{7.368,\,7.443,\,7.476\}$, while in the bottom panels we combine the absorption from all three absorbers and re-introduce the saturated Gunn-Peterson trough \citep{GP65} at $z<7.476$, beyond which the ionizing light from the quasar should be completely blocked. 

From the comparison in Figure~\ref{fig:absmodel}, it is clear that in the $[{\rm O}/{\rm H}]=-2.5$ case, the relevant portion of the quasar spectrum close to rest-frame Ly$\alpha$ -- i.e., within the Ly$\alpha$-transparent proximity zone and the immediate red-side continuum -- is almost entirely unaffected by the absorbers, thus we would expect that the resulting constraints on the IGM neutral fraction should be roughly the same. In the $[{\rm O}/{\rm H}]=-3.5$ case, however, it appears that both the drop-off of the proximity zone and the red-side damping wing shape can come almost entirely from the combined proximate absorption. 

Here we quantify the impact of the Ly$\alpha$ absorption by the three absorption systems on the IGM neutral fraction constraints presented in \citet{Davies18b} by repeating their analysis including the additional absorption by the foreground systems. We briefly summarize here their methodology. There are two primary components to the method: the determination of the intrinsic quasar continuum and its uncertainty, and the simulation-based inference of the IGM neutral fraction.

\begin{figure*}
\begin{center}
\resizebox{17.5cm}{!}{\includegraphics[trim={0.0em 1em 1.0em 1em},clip]{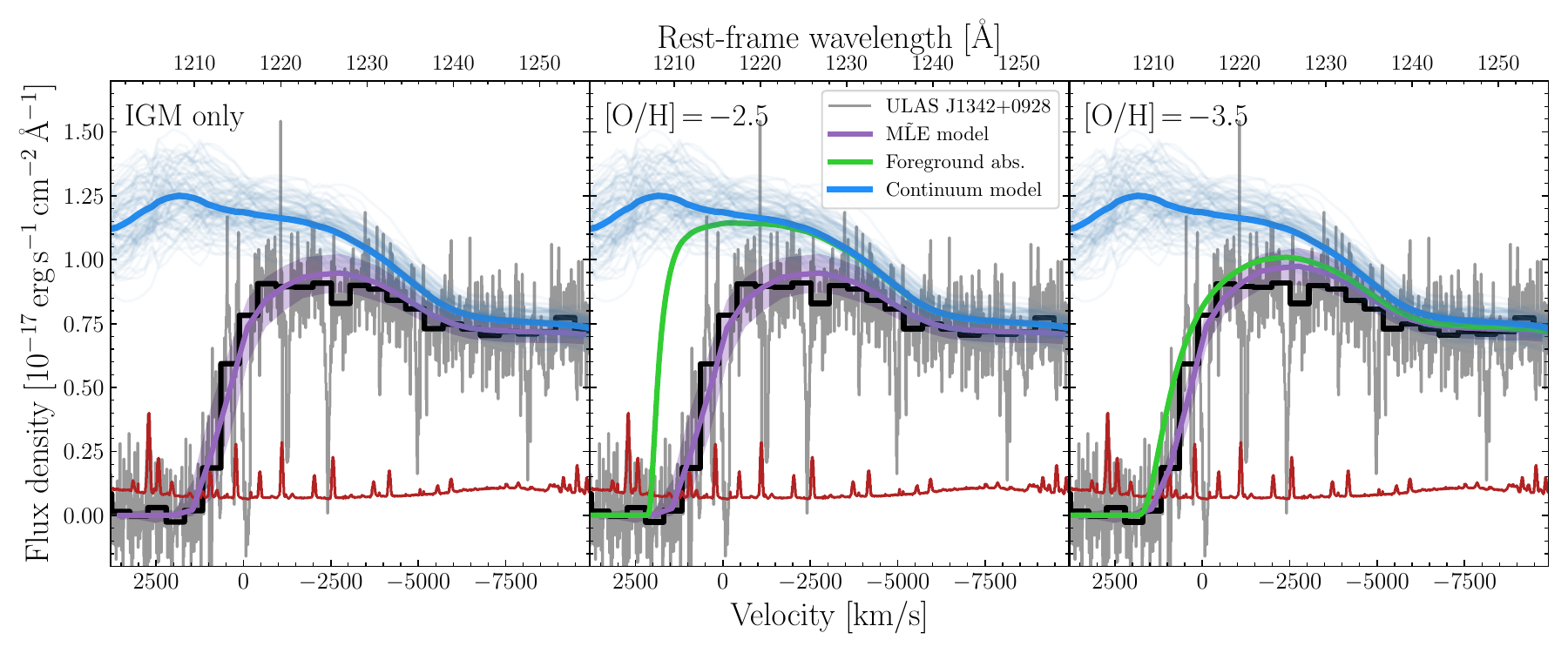}}
\end{center}
\caption{Best-fit absorption models in the absorber-free case with only IGM absorption (left; $\langle x_{\rm HI} \rangle=0.90$), the $[{\rm O}/{\rm H}]=-2.5$ case (middle; $\langle x_{\rm HI} \rangle=0.85$), and the $[{\rm O}/{\rm H}]=-3.5$ case (right; $\langle x_{\rm HI} \rangle=0.1$), where the latter two include both the effect of the foreground absorbers and the IGM. The purple curves and shaded regions show the median and 68\% scatter of the Ly$\alpha$ transmission of the maximum pseudo-likelihood model. The grey and black curves show the unbinned quasar spectrum and the 500\,km/s-binned spectrum, respectively, while the red curve shows the unbinned noise vector. The blue curve shows the prediction for the intrinsic continuum, with a swarm of transparent curves showing draws from its covariant uncertainty. The green curves in the middle and right panels show the corresponding model of the absorption from the three foreground absorbers. }
\label{fig:absmodel_fits}
\end{figure*}

The unabsorbed quasar continuum is estimated using the principal component analysis (PCA) approach described in \citet{Davies18a}. The ``training'' data consist of a sample of $12,764$ quasar spectra from BOSS DR12 with signal-to-noise ratio greater than 7 at rest-frame $1280$\,\AA. The principal component vectors are then derived from the logarithm of nearest-neighbor stacks of spline-smoothed spectra. Two separate sets of principal components are computed from ``red-side'' wavelengths ($1280<\lambda_{\rm rest}<2850$\,\AA) and ``blue-side'' wavelengths ($1170<\lambda_{\rm rest}<1280$\,\AA). For every quasar in the training set, the best-fit red-side and blue-side coefficients were fitted jointly with a redshift offset, introducing one additional non-linear degree of freedom that disconnects the spectra from the exact definition of the BOSS redshift determinations. A linear model (i.e.~a transformation matrix) is then fit to transform a set of best-fit red-side coefficients to a prediction for the blue-side coefficients. This procedure was then performed on every quasar in the training data to determine an empirical covariance matrix of the continuum prediction error. While some improvements to the quasar sample and training data have been made in the meantime \citep[e.g.][]{Bosman21, Chen22}, we use the original continuum prediction for J1342 as \citet{Davies18a} and \citet{Davies18b} for clarity and direct comparison.

We similarly use the same simulation data as \citet{Davies18b}. Skewers of density, temperature, and peculiar velocity are drawn through a high-resolution hydrodynamical simulation run with the Nyx code \citep{Almgren13,Lukic15}, with 4096$^3$ dark matter particles and baryon cells in a volume $100$\,Mpc$/h$ on a side. The initial ionization state of the gas is derived from a separate patchy reionization simulation run in a larger volume (256\,Mpc on a side) with a modified version of the 21cmFAST code \citep{Mesinger11,DF22}, where the overall ionizing efficiency of halos is adjusted to produce reionization topologies with volume-averaged IGM neutral fractions $\langle x_{\rm HI} \rangle$ in steps of 0.05 from 0 to 1. The stitched skewers are then fed into the one-dimensional ionizing radiative transfer (RT) code from \citet{Davies16} to compute the effect of the quasar radiation on the local IGM. We compute and save the corresponding Ly$\alpha$ transmission spectra in time-steps of quasar lifetime $\Delta\log{t_{\rm q}}=0.1$\footnote{Here we differ slightly from \citet{Davies18b} who used a coarser spacing, $\Delta\log{t_{\rm q}}=0.25$, instead opting for the finer spacing in quasar lifetime from \citet{Davies19}.} across a range from $t_{\rm q}=10^{3}$ to $10^8$\,yr.

To take the proximate absorbers into account, we add the DLA absorption directly to the RT model spectra in post-processing. That is, we do not re-run the RT calculations, but instead modify the output Ly$\alpha$ transmission spectra by adding the DLA optical depths to the original IGM-only optical depths at every pixel in velocity space. We further remove all transmitted flux at velocities beyond the highest redshift absorber (i.e.~$z_{\rm Ly\alpha}<7.476$), as the ionizing flux from the quasar should be fully attenuated. In this procedure we introduce a physical inconsistency: in the RT models, the neutral IGM beyond the first absorber will still be ionized away by the quasar. We expect that this effect is small, however, and should not impact our main results.

Finally, identically to \citet{Davies18b}, we employ a calibrated simulation-based inference approach to extract posterior constraints from the observed spectrum. The likelihood function of the quasar spectrum close to Ly$\alpha$ is extremely complicated, as it must include a variety of stochastic and covariant processes, thus precluding robust analysis using a multivariate Gaussian approximation. Instead, we identify a ``best-fit'' model in the coarse $21\times51$ space of $\{\langle x_{\rm HI} \rangle,t_{\rm q}\}$ via an approximate or ``pseudo''-likelihood, and use this as our summary statistic. The pseudo-likelihood is defined as the product of individual flux probability distribution functions (PDFs) of a 500\,km/s-binned representation of the spectrum. By running millions of mock observations, including the stochasticity in the small-scale structure of the IGM (via the hydrodynamical skewers) and the large-scale reionization topology (via the 21cmFAST skewers) as well as the covariant continuum prediction uncertainty, we then build a mapping between the best-fit parameter values and the true parameter values. In this way, we can translate a measurement of the best-fit parameters into a posterior PDF on the true parameters.

We recover maximum pseudo-likelihood parameter combinations of $\{\langle x_{\rm HI} \rangle,\log[{t_{\rm q}}/{\rm yr}]\}$) of $(0.90,6.2)$, $(0.85,6.2)$, and $(0.10,5.5)$ for the IGM only, $[{\rm O}/{\rm H}]=-2.5$, and $[{\rm O}/{\rm H}]=-3.5$ cases, respectively. In Figure~\ref{fig:absmodel_fits}, we show the 500\,km/s-binned spectrum of J1342 along with the corresponding median transmission curves and the 68\% scatter in the model, which for the middle and right panels includes the absorption by the foreground absorbers. The best-fit models are nearly identical, and all appear to be perfectly adequate representations of the quasar spectrum.

\begin{figure}
\begin{center}
\resizebox{8.5cm}{!}{\includegraphics[trim={0.0em 1em 0.0em 0em},clip]{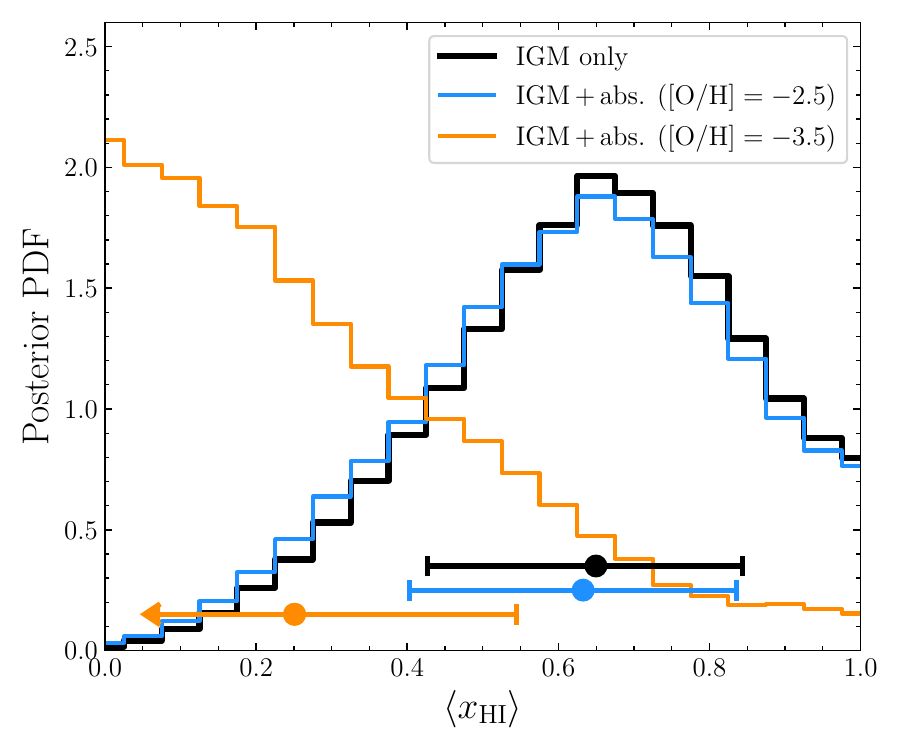}}
\end{center}
\caption{Posterior PDFs of the IGM neutral fraction $\langle x_{\rm HI} \rangle$ from the Ly$\alpha$ transmission analysis. The black curve shows the posterior assuming that the signal is entirely intergalactic, while the blue and orange curves include the presence of \ion{H}{1} absorbers corresponding to the $z>7.3$ low-ionization absorption systems found by \citet{Christensen23} assuming relatively high metallicity ($[{\rm O}/{\rm H}]=-2.5$) or low metallicity ($[{\rm O}/{\rm H}]=-3.5$). The points with error bars show the medians and central 68\% credible intervals; for the $[{\rm O}/{\rm H}]=-3.5$ case we elect to present them as an upper limit for clarity.}
\label{fig:xhipost}
\end{figure}

In Figure~\ref{fig:xhipost}, we show the resulting 1D posterior PDFs for the IGM neutral fraction $\langle x_{\rm HI} \rangle$ for the three cases, after marginalization over the quasar lifetime $t_{\rm q}$. The corresponding posterior medians and central 68\% credible intervals for $\langle x_{\rm HI} \rangle$ are $0.65^{+0.19}_{-0.22}$ and $0.63^{+0.20}_{-0.23}$ for the IGM only and $[{\rm O}/{\rm H}]=-2.5$ cases, respectively. For clarity, in the $[{\rm O}/{\rm H}]=-3.5$ case we instead quote 50\% and 84\% upper limits of $0.25$ and $0.54$, respectively. The relatively weak \ion{H}{1} absorption in the higher metallicity case has only a very minor impact on the IGM constraints, while the stronger \ion{H}{1} absorption in the lower metallicity case dramatically shifts the posterior PDF to lower values of $\langle x_{\rm HI} \rangle$. Notably, while the posterior PDF is in strong disagreement with a fully ionized IGM in the IGM-only case, in our model with strong \ion{H}{1} absorbers present the posterior PDF instead \emph{peaks} at an IGM neutral fraction of zero.

\section{Discussion \& Conclusion}\label{sec:disc}

In this work, we have investigated the implications of the discovery of weak low-ionization absorption systems within several thousand km/s from the quasar ULAS~J1342$+$0928 for its use as a probe of the neutral intergalactic medium during the epoch of reionization. From the equivalent widths of the \ion{Mg}{2} and \ion{O}{1} absorption lines, we infer plausible column densities for \ion{O}{1}, and convert these into \ion{H}{1} column densities assuming a range of oxygen abundances. For $[{\rm O}/{\rm H}]\lesssim-3$, the strongest absorbers reach high enough $N_{\rm HI}$, into the damped Ly$\alpha$ regime with $\log{N_{\rm HI}}>20.3$, to have an appreciable contribution to the absorption signal seen in the quasar spectrum. At a metallicity of $[{\rm O}/{\rm H}]=-3.5$, the proximate absorbers are strong enough to dominate the observed absorption signal, while much lower metallicites $[{\rm O}/{\rm H}]\lesssim-4$ would lead to stronger absorption than allowed by the observed spectrum.

We quantified the impact that putative absorbers could have on IGM neutral fraction constraints by repeating the damping wing analysis of \citet{Davies18b} with additional absorption injected into the simulated spectra. For a model of the absorbers with $[{\rm O}/{\rm H}]=-2.5$, the resulting change in the posterior PDF of the IGM neutral fraction is negligible, but for a model with $[{\rm O}/{\rm H}]=-3.5$, the IGM neutral fraction constraint shifts dramatically to lower values, instead preferring a completely ionized universe. We note that a DLA with such a low oxygen abundance has never been observed at any redshift (cf.~\citealt{Cooke17,Welsh23}). However, the decreasing level of ionization in metal absorbers at $z\gtrsim6$ (e.g.~\citealt{Cooper19, Becker19}) may transform gas clouds that would otherwise be a Lyman limit system (LLS) at lower redshifts into DLAs, and LLSs have been discovered with vanishingly small metallicities $\lesssim10^{-4}$ of solar \citep{Fumagalli11,Crighton16,Robert19}, so we cannot rule out this scenario \textit{a priori}.

\begin{figure}
\begin{center}
\resizebox{8.5cm}{!}{\includegraphics[trim={1.0em 1.25em 1.0em 0em},clip]{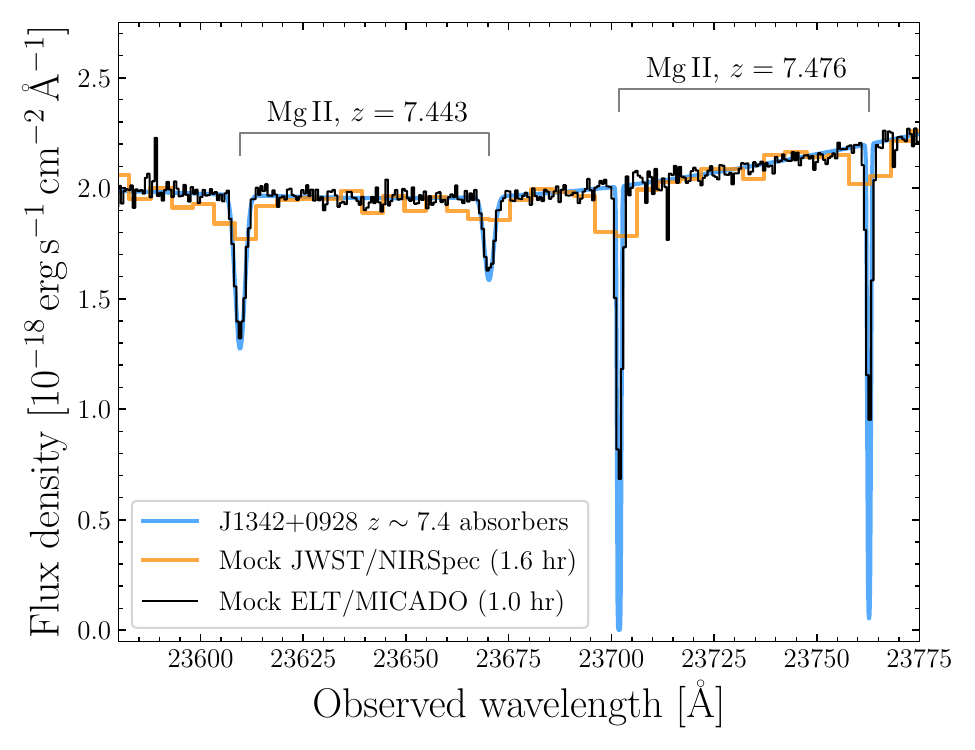}}
\end{center}
\caption{Segment of a mock J1342 spectrum centered on the $z=7.443$ and $z=7.476$ absorbers. The blue curve shows the assumed intrinsic model of the spectrum with the two absorbers using the physical parameters from Section~\ref{sec:abs}. The orange curve shows a mock JWST/NIRSpec G235H spectrum with resolution and signal-to-noise comparable to \citet{Christensen23}. The black curve shows a mock 1\,hr observation with ELT/MICADO at $R=20,000$.}
\label{fig:micado}
\end{figure}

We stress that there are still many unknowns in the analysis which preclude drawing any strong conclusions about the existence (or not) of the IGM damping wing in the J1342 spectrum. Despite our best efforts to account for saturation using the \ion{Mg}{2} doublet ratio, with only unresolved spectroscopy of the \ion{Mg}{2} and \ion{O}{1} absorption lines their corresponding column densities remain highly uncertain, and could be substantially higher than assumed here. With higher metal column density, the stringent requirements for low metallicity to produce DLA systems may be reduced. However, this uncertainty may be resolved in the near future via higher resolution spectroscopy from the next generation of ground-based telescopes, most notably the MICADO instrument \citep{Davies21MICADO} on the Extremely Large Telescope (ELT). In Figure~\ref{fig:micado}, we show a mock 1-hour ELT/MICADO spectrum\footnote{Using the ESO ELT spectroscopic exposure time calculator, \href{https://www.eso.org/observing/etc/}{https://www.eso.org/observing/etc/}, assuming a diffraction-limited observation in $K$-band with $R=20,000$.} of J1342 in the region surrounding the two highest redshift absorbers at $z=7.443$ and $z=7.476$, which achieves ${\rm S}/{\rm N}\sim80$ at a spectral resolution $R=20000$. For lack of direct access to the proprietary \citetalias{Christensen23} data, we compare the ELT/MICADO mock to a JWST/NIRSpec mock with resolution and signal-to-noise matched to the \citetalias{Christensen23} G235H spectrum. It is clear that ELT/MICADO will be extremely sensitive to weak \ion{Mg}{2} absorbers, and with sufficient spectral resolution to measure (or place much stronger constraints on) their kinematics. 

We have shown that the presence of absorbers like these have the potential to contribute additional systematic uncertainty in the reionization history derived from damping wing studies. In the context of future enlarged $z>7$ quasar samples in the era of near-IR space-based surveys like \textit{Euclid} and \textit{Roman}, deep JWST and ELT/MICADO observations will, therefore, be critical to interpreting damping wing observations. In addition to ``purifying'' samples of damping wing quasars via identifying metal absorbers within a few thousand km/s of systemic, the much larger search volume in the foreground of the quasars will allow for a sensitive survey of weak metal systems to determine their occurrence rate and thus the likelihood for any given quasar's proximity region to contain one. Forward modeling of this absorber population will be crucial to interpreting damping wing observations of large numbers of fainter quasars to be discovered by upcoming surveys (e.g.~\citealt{Barnett19}) for which such a sensitive absorber inventory is infeasible. 

Yet fainter objects like bright galaxies have been explored as a probe of reionization as well, now that JWST/NIRSpec is delivering continuum spectroscopy of large samples at $z>7$ (e.g.~\citealt{Umeda23}), but even the ELT will struggle to detect weak absorption lines in such objects. In addition, without the extreme ionizing flux from a nearby quasar, their local environments may be more susceptible to contamination by neutral dense (non-IGM) gas. This possibility has been hinted by the detection of DLAs intrinsic to some of the highest redshift galaxies, whose Ly$\alpha$ absorption is much stronger than could otherwise be imprinted by a fully neutral IGM \citep{Heintz23,Keating23}.

\section*{Acknowledgements}
We thank Jörg-Uwe Pott for discussions about ELT/MICADO capabilities.
S.E.I.B.~is supported by the Deutsche Forschungsgemeinschaft (DFG) through Emmy Noether grant number BO 5771/1-1.

\bibliographystyle{aasjournal}
 \newcommand{\noop}[1]{}

\end{document}